# Electrohydrodynamic settling of drop in uniform electric field at low and moderate Reynolds numbers

Nalinikanta Behera[1], Shubhadeep Mandal[2] and Suman Chakraborty[1,a]

[1]Department of Mechanical Engineering, Indian Institute of Technology Kharagpur, Kharagpur,West Bengal-721302, India

[2]Max Planck Institute for Dynamics and Self-Organization, Am Fassberg 17, D-37077 Göttingen, Germany

Dynamics of a liquid drop falling through a quiescent medium of another liquid is investigated in external uniform electric field. The electrohydrodynamics of a drop is governed by inherent deformability of the drop (defined by capillary number), the electric field strength (defined by Masson number) and the surface charge convection (quantified by electric Reynolds number). Surface charge convection generates nonlinearilty in a electrohydrodynamics problem by coupling the electric field and flow field. In Stokes limit, most existing theoretical models either considered weak charge convection or weak electric field to solve the problem. In the present work, gravitational settling of the drop is investigated analytically and numerically in Stokes limit considering significant electric field strength and surface charge convection. Drop deformation accurate upto higher order is calculated analytically in small deformation regime. Our theoretical results show excellent agreement with the numerical and shows improvement over previous theoretical models. For drops falling with moderate Reynolds number, the effect of Masson number on transient drop dynamics is studied for (i) perfect dielectric drop in perfect dielectric medium (ii) leaky dielectric drop in the leaky dielectric medium. For the latter case transient deformation and velocity obtained for significant charge convection is compared with that of absence in charge convection which is the novelty of our study. The present study suggests that for both the regimes, surface charge convection tends to increase or decrease the settling speed depending upon the ratios of electrical properties. Notably, in the inertial regime, deformation and velocity are seen to be altered prominently in the existence of significant charge convection.

## 1. Introduction

The dynamics of a drop moving through a continuous phase is extensively investigated owing to its inherent industrial applicability, which includes de-emulsification, drug delivery, ink-jet printing etc. (Anna 2016; Ghasemi *et al.* 2018; Mhatre *et al.* 2015; Teh *et al.* 2008) The most common application of gravity-induced motion of drop is oil-water phase separation process in oil industries (Berg *et al.* 2010; Eow *et al.* 2007; Mhatre *et al.* 2015). Due to the presence of tiny drops and low-density difference between oil and water, gravitational settling takes a long time. The separation efficiency can be enhanced by increasing the size of the drop thereby enabling it to traverse faster (Berg *et al.* 2010; Mhatre *et al.* 2015). The former can be achieved by electrocoalescence (Berg *et al.* 2010; Eow *et al.* 2001, 2007) of drops, while the drop settling speed can be increased significantly by applying external electric field (Mhatre *et al.* 2015).The need to understand the aforementioned phenomena is the motivation of the present work.

Application of electric field on a drop suspended in a continuous phase leads to some fascinating fluid dynmaics due to its ability to deform. After the pioneering work by Taylor (Taylor 1966), electrohydrodynamics (EHD) of drop got attention of many researchers. Taylor in his classical theory explained that no matter how small is the electrical conductivity, there will be accumulation of charges at the fluid-fluid interface. These surface charges along with the difference in electrical permittivity generates net normal and tangential components of electrical Maxwell stresses, which distorts the drop into either a prolate or oblate spheroid, depending on the ratios of electrical conductivity and permittivity. Following Taylor's theory, the EHD of drop and the influence of interfacial stresses have been studied extensively (Ajayi 1978; Lac & Homsy 2007; Melcher & Taylor 1969; Torza *et al.* 1971; Vizika & Saville 1992).The major limitations of the aforementioned theories are their negligence of charge convection. Later on taking charge convection into account, several studies (Das & Saintillan 2017; Feng 1999; Lanauze *et al.* 2015) have shown that prolate drops elongate more whereas oblate drops deform less in the presence of surface charge convection. It is to be noted that, most of the aforementioned studies have considered neutrally buoyant drops. For a translating drop, the deformation and speed of the drop were theoretically predicted by Spertell and Saville, wherein they have considered small drop deformation (Spertell & Saville 1974). In the limit of small deformation and small charge convection, Xu and Homsy (2006) presented corrected velocity and deformation for a settling drop through double asymptotic expansion analysis. While the drop is stationary, surface charge convection modifies the tangential electric stress, but the symmetric distribution is unaffected (Feng 1999). Hence drag on the drop is zero. However in case of settling drop, the asymmetric surface velocity breaks the symmetry in surface charge distribution and tangential electric stress (Bandopadhyay *et al.* 2016; Mandal *et al.* 2016b; Xu & Homsy 2006), hence affects the settling velocity. Although the theory proposed by Xu and Homsy (2006) shows qualitative agreement with their experimental results, it is to be noted that, their results lack a quantitative agreement. Furthermore, Yariv and Almong (Yariv & Almog 2016) for a weak electric field regime, have presented a theoretical model, depicting a non-monotonic behavior of the settling velocity, by extending the charge convection to a finite value. However in the limit of significant charge convection and electric field strength, the dynamics of drop settling is still unexplored.

Most of the theories discussed above explain the EHD of drop in Stokes flow regime (negligible inertial force compared to viscous force), where the drop remains spherical while moving through fluid medium. Taylor and Acrhivos (Taylor & Acrivos 1964) through singular perturbation analysis for low Reynolds number ($\text{Re}$, describes the relative strength of inertia force to viscous force) have presented that drops deform into oblate shape for the case of gravitational settling. In electric field, combined action of Maxwell stress and inertial stress can greatly influence the transient behavior of drop. Though in recent years the effect of electric field on bubble and drop dynamics has been studied by various researchers (Bararnia & Ganji 2013; Ghasemi *et al.* 2018; Wang *et al.* 2017; Yang *et al.* 2014) in inertia regime, the effect of charge convection is not mentioned.

In the present study, dynamics of gravity-driven drop subjected to uniform electric field has been investigated, for the case of Stokes regime ($\text{Re} \to 0$) and inertia regime (finite Re). For Stokes regime, the coupled EHD problem is solved theoretically, employing asymptotic analysis by considering small drop deformation in the presence of comparable electric field strength. Both drop and continuous phases are assumed to be poorly conducting (leaky dielectric) and the strength of charge convection is varied from a relative small to a finite value. A comparison between our analytical and numerical results has been demonstrated. For inertia dominated regime, transient drop dynamics is discussed based on our numerical results, considering two different drop-medium systems (i) perfect dielectric drop in perfect dielectric medium (ii) leaky dielectric drop in a leaky dielectric medium. For the second system, surface charge convection modulated transient deformation and settling velocity is presented considering significant strength of electric field.

## 2. Problem formulation

We consider a Newtonian drop of radius $a$, suspended in another Newtonian medium (refer to figure 1 for schematic representation). Fluid properties such as density, viscosity, electrical permittivity, and electrical conductivity are denoted by $\rho, \mu, \varepsilon, \sigma$. To distinguish drop phase and continuous phase, subscript '$i$' and '$e$' are used respectively. The difference in density assists the drop to settle under gravitational acceleration ($g\boldsymbol{e}_z$) with a constant velocity $\bar{U}_S$. Along with that an uniform electric field, $\boldsymbol{E_0} = -E_0\boldsymbol{e}_z$ is imposed opposite to the direction of gravity to study the concerning effect. The interfacial tension between the drop and continuous phase is assumed to be uniform and denoted by $\gamma$. The drop dynamics is assumed to be axisymmetric and a spherical coordinate system $(r,\theta)$ with its origin at the drop centroid is considered.

We non-dimensionalize different governing equations to describe the EHD problem by important dimensionless numbers. Throughout our study, the radius of the undeformed drop $(a)$ is used as length scale. Velocity is scaled by a reference velocity $\bar{U}_{ref}$. Following previous researchers (Bandopadhyay *et al.* 2016; Tripathi *et al.* 2015a, 2015b; Xu & Homsy 2006), we have chosen the reference velocity to be $\bar{U}_0$ and $\sqrt{ga}$ for Stokes regime and inertial regime respectively, where $\bar{U}_0 = \dfrac{2}{3}\dfrac{ga^2(\rho_i - \rho_e)}{\mu_e}\dfrac{(\mu_i + \mu_e)}{(3\mu_i + 2\mu_e)}$ is the Hadamard-Rybczynski velocity. The velocity scale is chosen such that, the correction to settling speed could be made noticeable due to application of electric field. The electric field is non-dimensionalized by the magnitude of external electric field $E_0$ and the surface charge density is scaled by $\varepsilon_e E_0$. The scales for hydrodynamic and electric stress used are $\dfrac{\mu_e \bar{U}_{ref}}{a}$ and $\varepsilon_e E_0^2$ respectively. Upon non-dimensionalization two dimensionless parameters arise viz. Masson number ($M = \dfrac{a\varepsilon_e E_0^2}{\mu_e \bar{U}_{ref}}$; describes the relative strength of electric stress compared to viscous stress) and electric Reynolds number ($Re_E = \dfrac{\varepsilon_e \bar{U}_{ref}}{\sigma_e a}$; the ratio of the charge relaxation time scale and flow time scale). The remaining dimensionless numbers and property ratios are stated below.

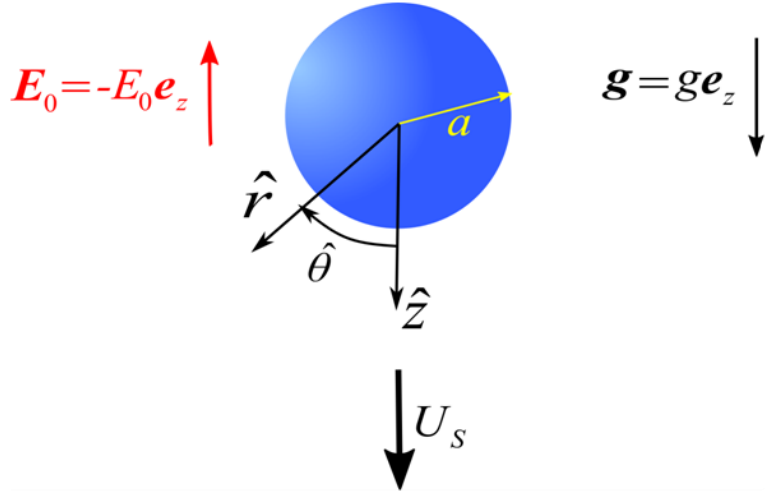

FIGURE 1. Schematic demostration of a viscous drop with radius $a$ settling under gravity $(g\bm{e}_z)$ with velocity $\bar{U}_S$ in unbounded domain and subjected to an uniform electric field $(\bm{E_0} = -E_0\bm{e}_z)$.

$$\mathrm{Re} = \frac{\rho_e \bar{U}_{ref} a}{\mu_e},\ Ca = \frac{\mu_e \bar{U}_{ref}}{\gamma},\ R = \frac{\sigma_i}{\sigma_e},\ S = \frac{\varepsilon_i}{\varepsilon_e},\ \lambda = \frac{\mu_i}{\mu_e}, \qquad (1)$$

where Re is the hydrodynamic Reynolds number, $Ca$ is the capillary number. $R, S, \lambda$ are the conductivity ratio, permittivity ratio, and viscosity ratio respectively.

## 2.1. *Governing equations and boundary conditions*

An EHD problem is governed by both the equations governing fluid flow and Maxwell's equations; coupled through a set of stress boundary conditions and charge conservation equation, which is to be discussed in the present section. As per leaky dielectric model, the electric potential ($\varphi$) inside and outside the drop satisfy the Laplace's equation (Bandopadhyay *et al.* 2016) as

$$\nabla^2 \varphi_i = 0, \quad \nabla^2 \varphi_e = 0. \tag{2}$$

The electric potential inside the drop must be bounded when $r \to 0$ and the electric potential outside the drop approaches the externally applied electric potential at far field i.e.

$$\varphi_i \text{ is bounded for } r \to 0, \tag{3}$$

$$\varphi_e \to \varphi_\infty = r\cos(\theta) \text{ at } r \to \infty. \tag{4}$$

Electric potential across the drop interface is continuous (Das & Saintillan 2017) which means

$$\varphi_i = \varphi_e \text{ at } r = r_S, \tag{5}$$

where $r_S = 1 + f(\theta)$ represents the deformed drop surface. Jump in electrical properties $(R, S)$ across interface produces Ohmic conduction, which is balanced by charge convection at the drop interface. The steady state, charge conservation equation in dimensionless form is written as (Mandal *et al.* 2017)

$$(E_{e,n} - RE_{i,n}) = -Re_E \nabla_\mathbf{s} \cdot (q_s \mathbf{V}_S), \tag{6}$$

where $E_n = \mathbf{n} \cdot \mathbf{E}$ represents the normal component of the electric field. The electric field can be calculated simply using $\mathbf{E} = -\nabla \varphi$. The unit normal vector $\mathbf{n} = \nabla(r - r_s)/|\nabla(r - r_s)|$ is acting outwards from the drop surface and $\nabla_s = (\mathbf{I} - \mathbf{nn}) \cdot \nabla$ is the surface gradient operator. In (6) $q_s$ represents the surface charge density which can be expressed as (Lac & Homsy 2007; Mandal *et al.* 2017)

$$q_s = E_{e,n} - SE_{i,n} \qquad (7)$$

and $V_S$ is the velocity along the interface. The net charge on drop surface must be zero (Yariv & Almog 2016),

$$\oint_{r=r_s} q_s dA = 0, \qquad (8)$$

where $dA = 2\pi r_S^2 \sin(\theta) d\theta$. It is important to note that in (6) presence of $V_S$ makes it nonlinear. To find the solution for $V_S$ it is necessary to solve the equations governing the fluid flow. Assuming incompressible flow, Navier-Stokes equation and continuity equation for Stokes flow simplify into

$$\lambda \nabla^2 \mathbf{u}_i = \nabla p_i, \quad \nabla \cdot \mathbf{u}_i = 0, \qquad (9)$$

$$\nabla^2 \mathbf{u}_e = \nabla p_e, \quad \nabla \cdot \mathbf{u}_e = 0, \qquad (10)$$

where $\mathbf{u} = u_r \mathbf{e}_r + u_\theta \mathbf{e}_\theta$ is the velocity field and $p$ is the pressure. Inside the drop, the velocity field is bounded and at far from the drop the flow is uniform (Kim *et al.* 2007; Mandal *et al.* 2016a) i.e.

$$\mathbf{u}_i \text{ is bounded as } r \to 0 \qquad (11)$$

$$\mathbf{u}_e \to -U_S \mathbf{e}_z \text{ as } r \to \infty \qquad (12)$$

Across the interface, normal velocities satisfy nopenetration boundary condition and the tangential velocities are continuous (Bandopadhyay *et al.* 2016):

$$\left. \begin{array}{c} u_{i,t} = u_{e,t} \\ u_{i,n} = u_{e,n} = 0 \end{array} \right\} \text{ at } r = r_S \qquad (13)$$

To get stable deformation, the forces acting due to viscous stress, electric stress and interfacial tension at the interface must be in equilibrium (Bandopadhyay *et al.* 2016; Mandal *et al.* 2017) i.e.

$$\boldsymbol{\tau}_e^H \cdot \mathbf{n} - \boldsymbol{\tau}_i^H \cdot \mathbf{n} + M\left(\boldsymbol{\tau}_e^E \cdot \mathbf{n} - \boldsymbol{\tau}_i^E \cdot \mathbf{n}\right) = \frac{1}{Ca}(\nabla \cdot \mathbf{n})\mathbf{n} \text{ at } r = r_s \qquad (14)$$

Where $\boldsymbol{\tau}^H$ is hydrodynamic stress tensor and $\boldsymbol{\tau}^E$ is electric stress tensor (Kim *et al.* 2007).

## 2.2. Small-deformation perturbation analysis

A closer look into (6) reveals that coupling of electric field with velocity field makes the EHD problem nonlinear. So to solve EHD problem of a drop settling under gravity, solutions of EHD problem of a neutrally buoyant drop in uniform electric field and hydrodynamics of gravitational settling drop can not superimposed. To solve such a problem analytically for arbitrary value of $Ca$, $Re_E$, $M$ is not an easy task. Therefore, we focus our study on a system where the drop is slightly deformed from initial spherical shape i.e. $Ca \ll 1$ keeping $Re_E \sim 1$ and $M \sim 1$, which refers to finite surface charge convection and electric field strength. Such choice of parameters is not arbitrary rather motivated by the experiments of Xu and Homsy (2006). They have conducted experiments on settling of PMM drop ($\rho_i = 1000\,\text{kg/m}^3$, $\mu_i = 0.5\,\text{Pa s}$, $\varepsilon_i = 2.8\varepsilon_0$, and $\sigma_i = 10^{-12}\,\text{S/m}$) in castrol oil medium ($\rho_e = 957\,\text{kg/m}^3$, $\mu_e = 1.4\,\text{Pa s}$, $\varepsilon_i = 4.45\varepsilon_0$ and $\sigma_i = 10^{-11}\,\text{S/m}$). Using these properties, considering $a = 3\,\text{mm}$ and electric field strength of $100\,\text{kV/m}$ various dimensionless parameters calculated are $Ca \sim 0.1$, $Re_E \sim 1$, $M \sim 1$, and $Re \sim 10^{-3}$. Very small value of $Ca$ allows us to employ asymptotic expansion using $Ca$ as perturbation parameter, following which any field variable $\zeta$, can be expressed as

$$\zeta = \zeta^{(0)} + Ca\,\zeta^{(Ca)} + Ca^2 \zeta^{(Ca^2)} + ..., \tag{15}$$

where '0' and '$Ca$' are used in the superscript to identify the leading-order and $O(Ca)$ contributions throughout our analytical study. In a similar way settling speed ($U_S$) and shape of drop ($r_s$) are expanded as (Mandal *et al.* 2016a)

$$U_s = U_S^{(0)} + CaU_S^{(Ca)} + O(Ca^2), \tag{16}$$

$$r_s = 1 + Ca\,f^{(Ca)} + Ca^2 f^{(Ca^2)} + O(Ca^3). \tag{17}$$

Here $f^{(Ca)}$ and $f^{(Ca^2)}$ are shape functions of order $O(Ca)$ and $O(Ca^2)$ respectively. The general form of these two for small drop deformation can be expressed as linear combination of Legendre polynomials

$$f^{(Ca)} = \sum_{n=0}^{\infty} \omega_n^{(Ca)} P_n(\eta)\,, \quad f^{(Ca^2)} = \sum_{n=0}^{\infty} \omega_n^{(Ca^2)} P_n(\eta). \tag{18}$$

Here $P_n(\eta)$ is the Legendre polynomial of order $n$ and $\eta = \cos(\theta)$.

## 3. Solution procedure

In this section, we discuss the analytical procedure that we have implemented to achieve the final solution. As the electric potential inside and outside of the drop satisfies the Laplace equation, the general expression for $\varphi_i$ and $\varphi_e$ are

$$\varphi_i = \sum_{n=0}^{\infty} \left[ \bar{\alpha}_n r^n + \bar{\beta}_n r^{-(n+1)} \right] P_n(\eta),$$
$$\varphi_e = \sum_{n=0}^{\infty} \left[ \alpha_n r^n + \beta_n r^{-(n+1)} \right] P_n(\eta). \tag{19}$$

Using boundary condition (3) and (4) the electric potentials can be modified into

$$\varphi_i = \sum_{n=0}^{\infty} \bar{\alpha}_n r^n P_n(\eta),$$
$$\varphi_e = r\cos(\theta) + \sum_{n=0}^{\infty} \beta_n r^{-(n+1)} P_n(\eta). \tag{20}$$

Now using stream functions $\psi_i$, $\psi_e$ instead of respective velocity fields, (9) and (10) can be transformed into

$$\Omega^2\left(\Omega^2 \psi_i\right) = 0, \quad \Omega^2\left(\Omega^2 \psi_e\right) = 0, \tag{21}$$

where $\Omega^2 = \dfrac{\partial^2}{\partial r^2} + \dfrac{1-\eta^2}{r^2}\dfrac{\partial^2}{\partial \eta^2}$. The general solution to (21) is of the form (Mandal *et al.* 2017)

$$\psi_i = \sum_{n=1}^{\infty} \left( \bar{A}_n r^{n+3} + \bar{B}_n r^{n+1} \right) Q_n(\eta),$$
$$\psi_e = U_S r^2 Q_1(\eta) + \sum_{n=1}^{\infty} \left( C_n r^{2-n} + D_n r^{-n} \right) Q_n(\eta). \tag{22}$$

Here $Q_n(\eta) = \int_{-1}^{\eta} P_n(\eta) d\eta$ is the Gegenbauer polynomial of degree $n$. Expanding the field variables using (15) and substituting the same into the equations provided in section 2.1, we obtain and $O(Ca)$ governing equations and different boundary conditions. Along with these equations one need to consider force balance on the drop to get closed system of equations, which can be shown by

$$\boldsymbol{F}_B + \boldsymbol{F}_H + M\boldsymbol{F}_E = 0, \tag{23}$$

Where $\boldsymbol{F}_B = 2\pi\left(\dfrac{3\lambda+2}{\lambda+1}\right)\boldsymbol{e}_z$ is the buoyancy force, $\boldsymbol{F}_H = 2\pi \int_0^{\theta=\pi} \left(\boldsymbol{\tau}_e^H \cdot \boldsymbol{n}\right) r_s^2 \sin\theta d\theta$ is the hydrodynamic force and $\boldsymbol{F}_E = 2\pi \int_0^{\theta=\pi} \left(\boldsymbol{\tau}_e^E \cdot \boldsymbol{n}\right) r_s^2 \sin\theta d\theta$ is the electric force. The detailed analytical treatment is provided in the supplementary material. Now we should mention the important assumption made to reach the final solution. Equations (28) and (29) suggest that for nonzero $Re_E$, both the electric field and velocity field contains infinite number of nonzero spherical harmonics. Solving for large number of spherical harmonics is not an easy task. To make progress, we have followed the small-deformation theory developed by Das and Saintillan (2017). We have considered only first 5 terms in series of harmonics for leading-order solutions and 10 spherical harmonics for $O(Ca)$. The choice of the number of terms is such that, by increasing the number of harmonics the drop velocity is almost unaffected ( please refer to appendix-A). However the higher order harmonics get stronger with increasing $Re_E$, which constraints the present theory for systems having low and moderate electric Reynolds number (Das & Saintillan 2017). Along with velocity charge convection also affects the shape deformation of the drop (Feng 1999; Xu & Homsy 2006). The shape deformation of the drop is characterized by the deformation parameter ($D$) which is calculated by

$$D = \frac{l^{\parallel} - l^{\perp}}{l^{\parallel} + l^{\perp}}, \qquad (24)$$

where $l^{\parallel}$ and $l^{\perp}$ are the length of the drop parallel and perpendicular to electric field respectively. $D>0$, indicates prolate deformation whereas oblate deformation is defined by $D<0$. As mentioned previously we calculate deformation up to higher order, given as

$$D = CaD^{(Ca)} + Ca^2 D^{(Ca^2)} + O(Ca^3) \qquad (25)$$

## 4. Numerical simulation

The theoritical model presented above is able to predict the shape deformation and drop speed only when the deformation is samll and the flow is in the limit of vanishingly small Reynolds number. However to explore the regime of finite drop deformation and finite fluid inertia, we perform numerical analysis. We use an open source fluid flow solver, Gerris developed by Popinet (Popinet 2003, 2009) to simulate the present problem. Assuming incompressible flow, Navier-Stokes equation is solved using finite volume method along with VOF (volume of fluid) method to track the interface accurately. Dynamic adaptive grid refinement is incorporated in Gerris, which allows us to use very small grid size near the interface and relatively larger grid size far from it, hence optimizes the computational time. Besides the flow problem, it is also very useful to solve EHD problems (López-Herrera *et al.*

2011). The non-dimensionalized form of continuity and momentum conservation equation used are as follows

$$\nabla \cdot \boldsymbol{u} = 0, \tag{26}$$

$$\mathrm{Re}\left(\frac{\partial \boldsymbol{u}}{\partial t} + \boldsymbol{u} \cdot \nabla \boldsymbol{u}\right) = -\frac{M}{\rho}\nabla p + \frac{\mu}{\rho}\nabla^2 \boldsymbol{u} + \frac{\boldsymbol{F}_\gamma}{\rho Ca} + M\boldsymbol{F}_E + \frac{\mathrm{Re}}{Fr}\boldsymbol{e}_z. \tag{27}$$

In the above equation $F_\gamma = \delta \kappa \boldsymbol{n}$ and $F_E = q_v \boldsymbol{E} - (1/2)E^2 \nabla \varepsilon$ are the surface tension and electric force per unit volume respectively. $Fr = \bar{U}_{ref}^2 / ga$ is the Froude number whereas other dimensionless quantities are discussed in section 2. To show the effect of electric field, the pressure is scaled by $\varepsilon E_0^2$. The volume fraction ($c$) satisfies the advection equation

$$\frac{\partial c}{\partial t} + \boldsymbol{u} \cdot \nabla c = 0. \tag{28}$$

Dimensionless density, viscosity, permittivity, conductivity of the fluids are calculated using weighted average mean (WAM) method (López-Herrera *et al.* 2011) as

$$\begin{aligned}
\rho &= c\rho_r + (1-c), \\
\mu &= c\lambda + (1-c), \\
\varepsilon &= cS + (1-c), \\
\sigma &= cR + (1-c).
\end{aligned} \tag{29}$$

Here $\rho_r$ is the density ratio and other parameters are defined in section 2. For drop phase $c=1$ and for bulk fluid $c=0$ is considered. Using various scales discussed in section 2, the charge convection equation transforms into (López-Herrera *et al.* 2011)

$$Re_E \left(\frac{\partial q_v}{\partial t} + \nabla \cdot (q_v \boldsymbol{u})\right) = \nabla \cdot (\sigma \nabla \varphi), \tag{30}$$

where $q_v$ is the volumetric charge density. The electric potential satisfies Poissons equation as

$$\nabla \cdot (\varepsilon \nabla \varphi) = -q_v. \tag{31}$$

In the present study for both Stokes regime and inertial regime flow, an axisymmetric 2D domain of height $H$ and width $W$ is considered as shown in figure 2. The drop dynamics is assumed to be symmetric about the axis (shown by dotted vertical line). Neumann boundary condition for flow field is imposed at all other sides of the domain. Except axis, $\varphi = -E_0 z$ is specified at each boundary to employ upward uniform electric field. To perform simulation in Stokes flow, a domain of size $70a \times 35a$ is considered and while considering inertial effects a domain of size $60a \times 20a$ is considered. Initially, the drop is placed at a vertical location, $z_c \sim 10a - 22a$ from the top boundary so that it has no effect on the drop. To capture the interface accurately a cell size of $\Delta r = \Delta z = W \times 2^{-11}$ is taken near the interface whereas bulk mesh size is kept at $\Delta r = \Delta z = W \times 2^{-6}$. We have validated the numerical code for the (i) gravitational settling in the absence of electric field with Hadamard-Rybczynski theory (ii) electric field modified deformation with numerical results of Lac and Homsy (Lac & Homsy 2007). Grid independence and domain independence tests are performed. The details of validation of present numerical formulation is provided in appendix-B. The grid and domain independence study are provided in appendix-C. While settling in Stokes regime in absence of electric field, dimensionless drop speed obtained numerically is $U_S^{E=0} \sim 0.9845$ (refer to figure B1(a)) instead of 1.0. While electric field is present, to avoid such numerical error, final velocity is calculated using $U_S = U_S / U_S^{E=0}$.

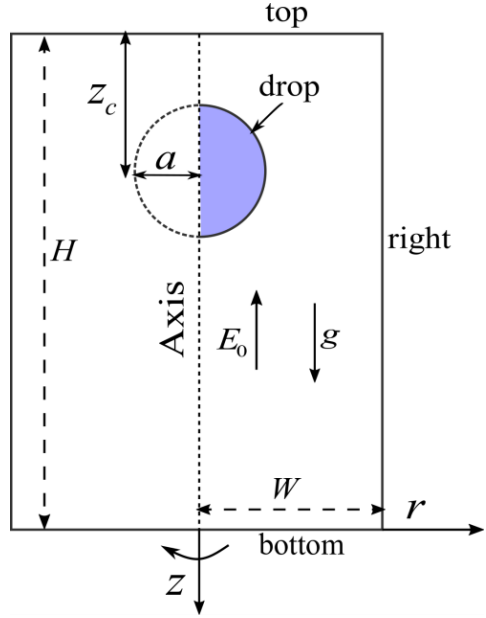

FIGURE 2. Problem setup for numerical simulation. Axisymmetric drop of radius $a$ is placed at a vertical distance of $z_c$ from top surface in a domain of size $H \times W$. Uniform electric field $(E_0)$ is applied in upward direction and drop is settling under gravitational acceleration $(\boldsymbol{g} = g\boldsymbol{e}_z)$.

# 6. Results and discussion

## 6.1. *Stokes regime*

In the present section combined influence of viscous force and electric force on drop dynamics is studied. To show the practical significance of our theory, the experimental data from Xu and Homsy (2006) (drop phase is PMM, $a = 3.5\,mm$, $\rho_i = 1000\,kg/m^3$, $\mu_i = 0.5\,Pa\,s$ and medium phase is castrol oil, $\rho_e = 957\,kg/m^3$, $\mu_e = 1.4\,Pa\,s$) are adopted, from which we obtain $\text{Re} \sim 2.4 \times 10^{-3}$, $Fr \sim 2.82 \times 10^{-5}$, which is a case of Stokes flow. Two sets of electrical properties $(R,S)$ are chosen for studying the effects of $Ca$. For one set of $R,S$ drop undergoes prolate deformation ($\phi_T > 0$), for another drop deforms into an oblate shape ($\phi_T < 0$), where $\phi_T = 1 + R^2 - 2S + 3(R-S)\dfrac{(3\lambda+2)}{5(\lambda+1)}$, is Taylor's discriminating parameter.

### 6.1.1. *Effect of capillary number*

In this section, we study the effect of capillary number $(Ca)$, which signifies the relative strength of viscous force to surface tension force on drop deformation and resulting settling speed. Surface charge convection (defined by $Re_E$) is considered and kept small for this case in order to compare our results with the asymptotic solution (Xu & Homsy 2006). Analytical results are provided only when deformation is small (for small values of $Ca$), while numerical simulations results are provided for higher values of $Ca$. Figure 5(a) depicts the deformation $(D)$ vs. $Ca$ plot for a drop undergoing oblate deformation i.e. $\phi_T < 0$, for material properties

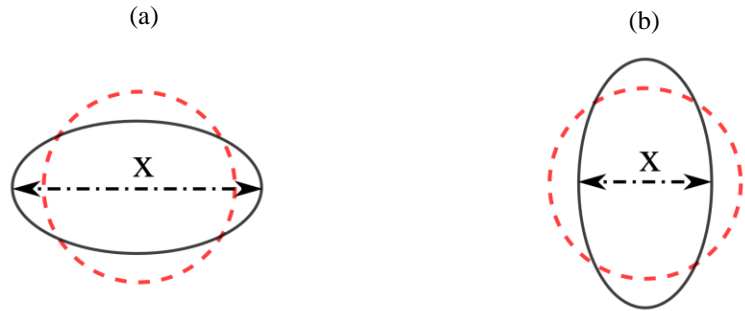

FIGURE 3. Schematic showing case of (a) oblate drop deformation (b) prolate drop deformation. Red dashed line represents the spherical drop and black solid line represents the deformed drop. The cross-sectional area is $A_c = \pi x^2/4$, where x is the width of the deformed drop.

$R = 1, S = 3, \lambda = 0.3571$ considering $M = 1$ and $Re_E = 0.2$. By increasing $Ca$, the relative strength of electric field as compared to surface tension force, defined by electric capillary number ($Ca_E = Ca \cdot M$) increases, thus allows the drop to deform more. By taking a closer look into figure 4(a) make it clear that our higher order theory shows a moderate improvement over asymptotic theory and predicts the deformation quite closer to the numerical results. $O(Ca^2)$ shape functions are not calculated in asymptotic solution of Xu and Homsy (2006), therefore shows discrepancy. The flow obstruction caused by an oblate spheroid owing to increase in projected area $(A_c)$ (refer to figure 3a) augments the hydrodynamic drag on it, hence results fall in drop speed. The drop settling speed $\left(U_S = U_S^{(0)} + Ca U_S^{(Ca)}\right)$ reduces continuously with increase in $Ca$, displayed in figure 4(b). The velocity calculated by the nonlinear theory matches precisely with the numerically obtained results. Undoubtedly for a very small deformation (small $Ca$), the asymptotic theory provides suitable results, however at an elevated deformation shows significant departure, despite the fact that weak charge convection is considered. The reason behind the inconsistency is that the asymptotic theory fails to explain $O(Ca)$ charge convection (S14) and $O\left(Re_E^2\right)$ contributions are not taken into account.

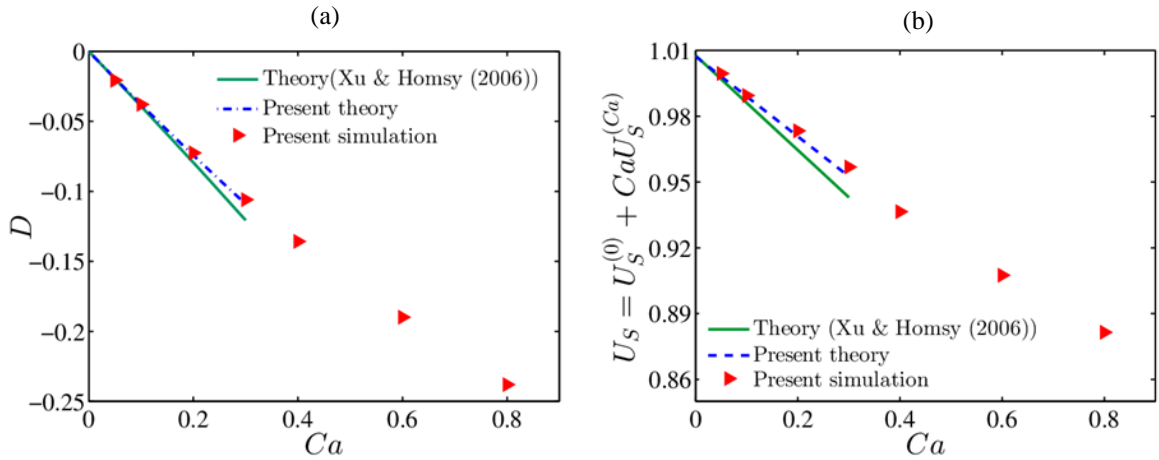

FIGURE 4. (a) Deformation parameter $D$ as a function of $Ca$ (b) Variation of settling speed with $Ca$ for $M = 1$, $Re_E = 0.2$. Material properties considered are $R = 1, S = 3, \lambda = 0.3571$.

Now we investigate a different case, where electric field assists prolate deformation. Figure 5(a) shows the effect $Ca$ on drop deformation $(D)$ for material properties $R = 3, S = 1, \lambda = 0.3571$, considering $M = 1$ and $Re_E = 0.2$. This is clearly a case of prolate deformation ($\phi_T > 0$), i.e. the drop tends to elongate in the direction of the applied electric field. Almost linear increase in drop deformation is observed with increase in $Ca$. On increasing $Ca$, the drop deformability becomes more, hence allowing the electric stress to distort the drop into more prolate spheroid shape. From figure 5(a), it is evident that our higher order theory predicts accurate deformation that of numerical results and outperforms the asymptotic theory. In general prolate deformation associated with drag reduction due to less projected area $(A_c)$ as shown in the schematic (figure 3b), hence allows the drop to move faster. Figure 5(b) shows the variation of settling velocity with $Ca$, considering above discussed parameters. Results obtained by present theory agrees well with the numerical results, however, fails to show any substantial improvement over the asymptotic theory. The reason behind this insignificant deviation is that, the prolate drop is weakly affected by charge convection, as reported previously by Das and Saintillan (2017). The deformation predicted by their nonlinear theory only able to show modest improvement over Taylor's asymptotic theory. When the drop is spherical ($Ca = 0$), the value of settling speed is seen to be slightly below 1 which is a consequence of surface charge convection, that is to be discussed in section 6.1.2.

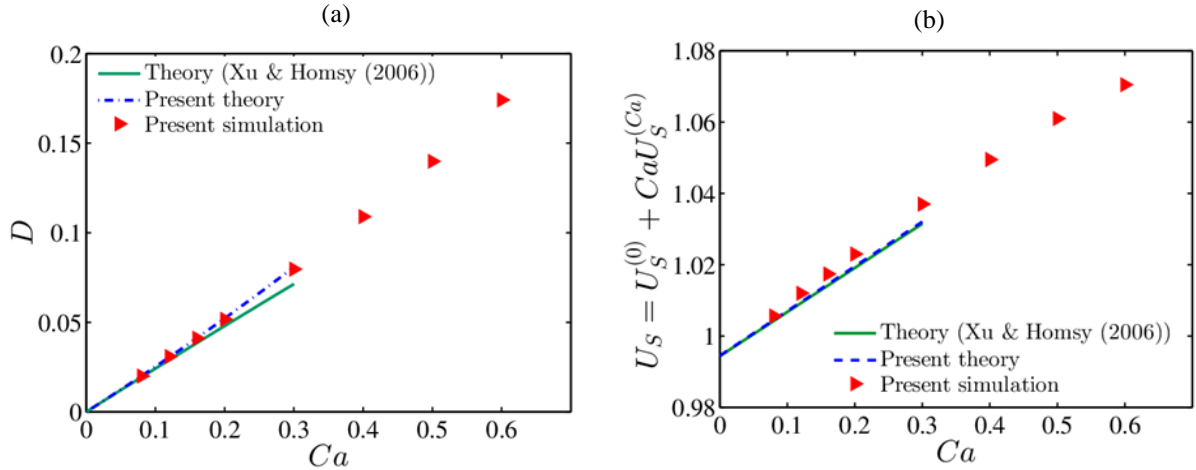

FIGURE 5. (a) Deformation parameter $D$ as a function of $Ca$ (b) Variation of settling speed with $Ca$ for $M = 1$, $Re_E = 0.2$. Different parameters considered are $R = 3, S = 1, \lambda = 0.3571$.

6.1.2. *Effect of nonlinear charge convection*

Now we focus our attention on the surface charge convection (strength of which is defined by $Re_E$) controlled drop dynamics considering negligible drop deformation (corresponds to $Ca \ll 1$). At small $Re_E$, the correction to settling speed owing to charge convection is characterized by a discriminating function, $\chi = (R - S)(S - 3R - 3)$ (Xu & Homsy 2006; Yariv & Almog 2016). While $\chi < 0$ is a sign of drop retardation, for $\chi > 0$, drop travels faster. Figure 6(a) shows the variation of settling speed ($U_S$) with $Re_E$ for $R = 2.5$, $S = 1$,

$\lambda = 0.3571$, $Ca = 0.05$, $M = 0.8$. Since in this case $\chi < 0$, velocity of the drop decreases continuously as the surface charge convection gets intensified. The present theory ables to predict the nonlinear variation of drop velocity quite well and shows a decent agreement with the numerical. For non-zero $Re_E$, the deviation in $U_S$ is a consequence of modified tangential electric stress distribution at drop interface as mentioned by previous researchers (Bandopadhyay *et al.* 2016; Xu & Homsy 2006). To describe this phenomenon briefly, first we plot the distribution of leading-order surface charge density ($q_s^{(0)}$) along drop interface in figure 6(b) for different $Re_E$. For $Re_E = 0$ (corresponds to absence of charge convection), charge distribution is anti-symmetric about the equator ($\theta = 90^0$). However for a translating drop, when surface charge convection is considered, spherical harmonics of order 2,3,4,5… come into picture, which are absent while $Re_E = 0$. This breaks the anti-symmetry in charge distribution making it asymmetric. With the increase in $Re_E$ these coefficients get enhanced and so does the asymmetry as shown in the figure 6(b). It can be seen that with an increase in $Re_E$, the surface charge density at the front end ($\theta = 0^0$) decreases whereas at the rear end ($\theta = 180^0$) the surface charge density decreases. For the present choice of electric properties ($R, S$), the secondary rolls produced because of the electric field are not noticeable. Hence the charges get transported by the pole to pole fluid circulation inside the drop induced by the motion of the drop, which is reflected in figure 6(b).

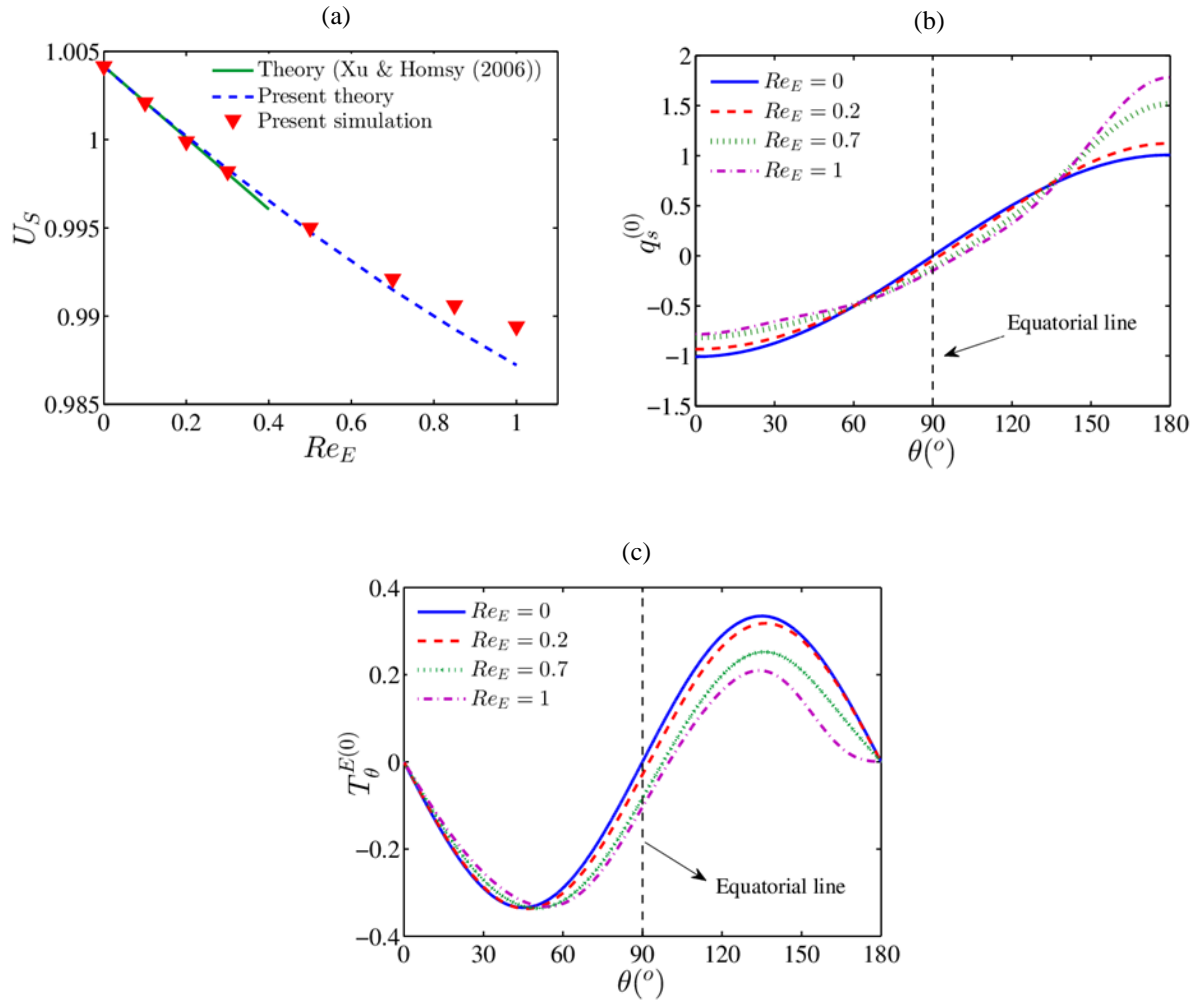

FIGURE 6. (a) Variation of settling speed with $Re_E$ (b) leading-order surface charge distribution obtained from present theory (c) tangential electric stress variation for $Ca = 0.05, M = 0.8$. Material properties considered are $R = 2.5$, $S = 1$, $\lambda = 0.3571$.

Figure 6(c) shows the distribution of leading-order tangential electric stress, $T_t^{E(0)} = q_s^{(0)} E_\theta^{(0)}$ along with the interface for $Re_E$. For $Re_E = 0$, the tangential electric stress produced is anti-symmetric about equator ($\theta = 90^0$). However, as the charges at the interface reallocated by charge convection, the tangential electric stress distributed along the interface is asymmetric about the equator. To balance the asymmetric tangential electric stress, flow pattern near the interface is modified to produce asymmetric hydrodynamic tangential stress. From figure 6(c) it is evident that, by increasing $Re_E$ the asymmetry in $T_t^{E(0)}$ further improved, hence alters the drag on the drop.

Figure 7(a) shows the variation of settling speed ($U_S$) with $Re_E$ for $R = 1, S = 2.5$, $\lambda = 0.3571$, $Ca = 0.05$, $M = 0.8$ which suggests $\chi > 0$. For the above properties, drop speed increases monotonically, when surface charge convection is taken into account. As already discussed in section 6.1.1, the velocity predicted by asymptotic theory shows poor prediction for settling speed even for small $Re_E$. The drop deformation is considered to be negligible ($Ca = 0.05$); consequently, the asymptotic theory matches well with the present nonlinear theory. The nonlinearity associated with the variation in settling speed is well captured by present theory; however, deviates slightly from the numerical for comparatively strong charge convection. The imprecision is attributed to the fact that in leading-order, only 5 number of spherical harmonics are considered from an infinite series, as already discussed in section 3. For the above case leading-order charge distribution is plotted along drop interface in the figure 7(b) for different $Re_E$. From the same figure, it is noticed that asymmetry generated by charge convection is more intense for $\chi > 0$. As a consequence, at both the poles ($\theta = 0^0$ and $\theta = 180^0$) a substantial deviation in surface charge density is observed. Alike surface charge density, modulation to strength and symmetry of electric tangential stress is affected significantly as seen in figure 7(c). Comparing the case of $Re_E = 0$ and $Re_E = 1$, near the front end of the drop, the strength of tangential electric stress reduces significantly while near the opposite end, the point of maximum stress shifted towards $\theta = 180^0$ with a quite higher strength.

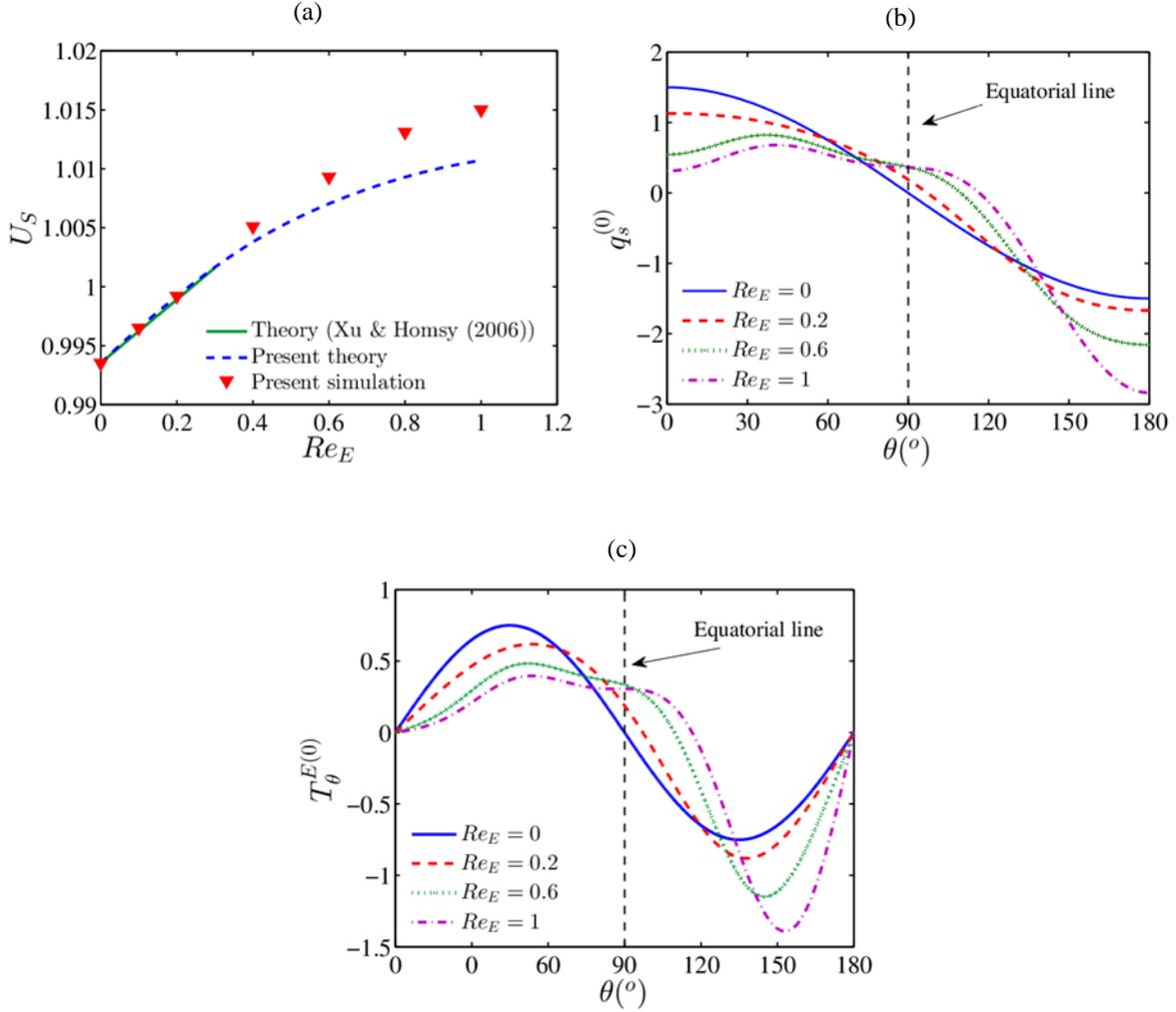

FIGURE 7. (a) Variation of settling speed with $Re_E$ (b) leading-order surface charge distribution obtained from present theory (c) tangential electric stress variation for $Ca = 0.05$, $M = 0.8$. Material properties considered are $R = 1$, $S = 2.5$, $\lambda = 0.3571$.

### 6.2. *Inertial regime*

Now we discuss the combined effect of gravitational settling and uniform electric field on a suspended drop in a flow regime where inertia force is relatively larger than viscous force i.e. case of finite Re. Drops moving through a quiescent medium with appreciable inertia generally deforms into an oblate shape reported by several researchers. To show the practical relevance of the present numerical study, we adopt the physical properties from the experiment conducted by Hu and Kintner (Hu & Kintner 1955). For a tetrachloroethylene drop falling through water medium, the density ratio ($\rho_r$) and viscosity ratio ($\lambda$) obtained are ~1.6192 and ~1.0 respectively. To illustrate the effect of electric field strength (defined by $M$) on drop deformation and velocity alternation, we have considered two drop-suspending media systems:

(i) perfect dielectric- perfect dielectric (PD-PD) & (ii) leaky dielectric-leaky dielectric (LD-LD). To avoid any instability in drop dynamics triggering by inertial forces, values of Re up to $O(10)$ is considered. It is assumed that the deformation in this regime is axisymmetric around z-axis and settling path is linear (Tripathi *et al.* 2015a). For a LD-LD system, moderately strong surface charge convection is considered to study its effect on drop dynamics.

6.2.1. *Perfect dielectric drop in a perfect dielectric medium*

In figure 8(a) we have shown the comparison between the deformation obtained for a settling drop and a neutrally buoyant drop for different values of $M$ considering $S = 5$, $Ca = 0.2$, $\lambda = 1$, $\text{Re} = 15$. For a difference in permittivity (value of $S$ other than 1), a neutrally buoyant PD drop always elongates in the direction of the applied electric field (Sherwood 1988). However, a settling drop can attain either a steady state prolate or oblate shape in finite inertial regime depending upon the electric field strength. While inertia stress promotes oblate deformation, electric stress tries to elongate the drop. Therefore the steady-state shape of the drop is determined by the relative strength of electric stress and inertia stress i.e. $\frac{M}{\text{Re}}$. With the increase in $M$, electric stress gets stronger against inertia stress, which is reflected in a gradual decline in oblate deformation; more spherical drop shape is obtained. For the above considered parameters, it has been found that for $M = 6.5$ (which gives $\frac{M}{\text{Re}} \sim 0.433$), electric stress is sufficient to balance the inertia stress and as an outcome, the shape of the drop becomes approximately spherical ($D \sim 10^{-4}$). Subsequently, increase in $M$ enhances the prolate deformation. For better understanding we have shown a comparison between the steady drop shapes of a neutrally buoyant drop and a settling drop at different values of $M$ in figure 8(b). Figure 8(c) depicts deformation vs. time plot for $M = 0, 3, 6$. In the absence of an electric field ($M = 0$), drop falling under gravity deforms into oblate shape monotonically with time. Similarly, a monotonic prolate deformation occurs owing to the sole effect of electric field, however presence of both brings about the non-monotonic transient behavior in deformation that can be seen in figure 8(c). For $M = 3$ and $M = 6$, it is observed that initially the drop deforms continuously into an asymmetric prolate shaped spheroid ($D > 0$) for $t < t_{MD}$, where $t_{MD}$ is the time required to achieve maximum deformation. For $M = 3, 6$ the corresponding $t_{MD} = 3, 4$ respectively. Initially the drop is slowly gaining the motion, hence in the time region $t < t_{MD}$, the deformation is mostly affected by electric stress rather than inertia generated stress, as a consequence of which we can see that the deformation obtained at $t=t_{MD}$ is similar to the steady-state deformation value for a neutrally buoyant drop for equal value of $M$. At $t=t_{ss}$, the drop attains steady-state deformation. For all the cases, $t_{ss} \sim 25$. In the time region $t_{MD} < t < t_{ss}$, inertia force is quite significant, as a result, deformation monotonically relaxes to a a steady-state value. For $M=3$ and $M=6$ the strength electric stress is not enough to balance the inertial stress, therefore the drop finally remained to be in oblate shape. As increasing $M$ decreases the cross-sectional area, the drop speed rises as represented in figure 8(d). It has been observed that by applying an electric field, a significant rise in drop velocity can be achieved. In the absence of an electric field ($M = 0$), the velocity of the drop is ~0.8821 whereas for $M = 6$, the velocity obtained is ~1.0671.

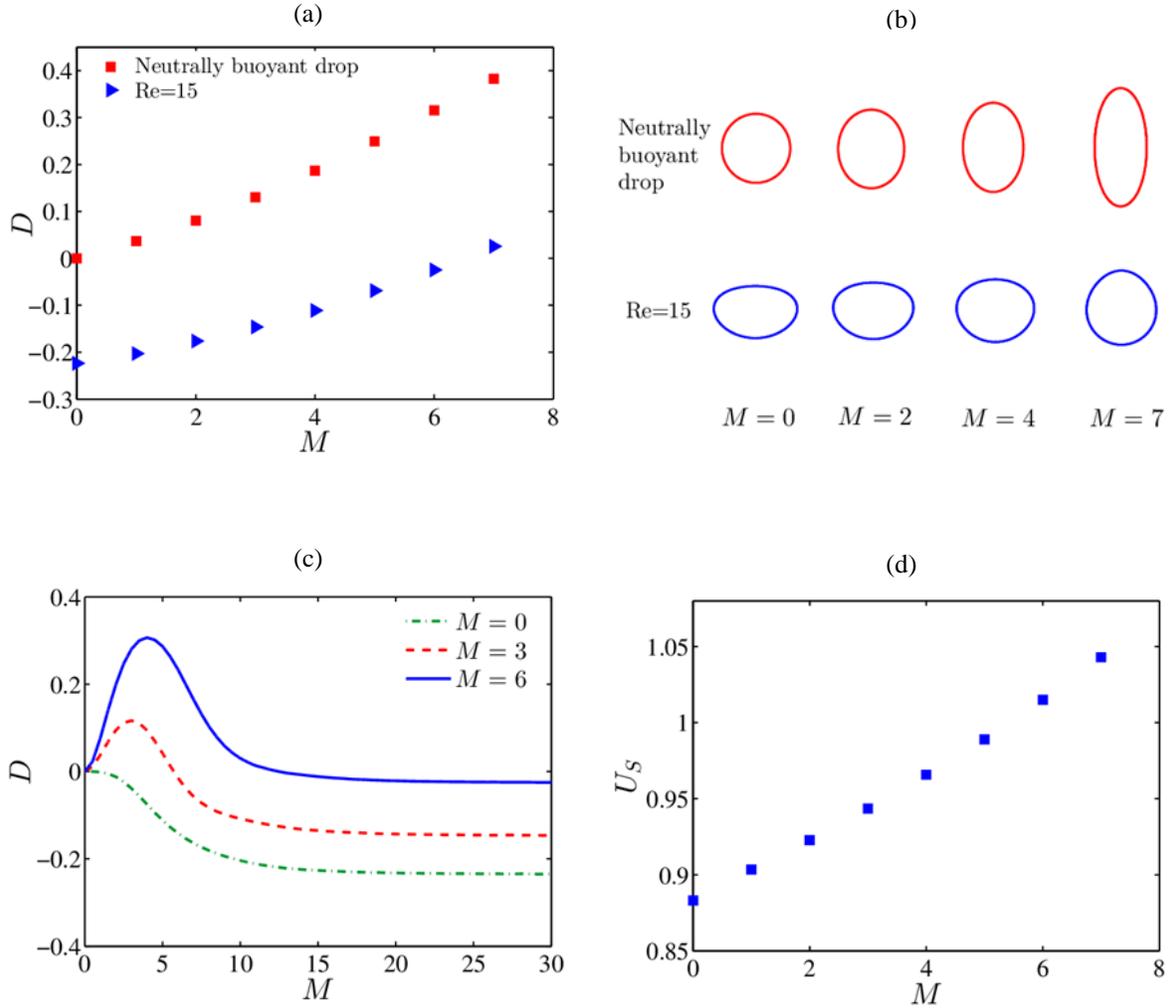

FIGURE 8. (a) Deformation ($D$) at different values of $M$ (b) Comparison between neutrally buoyant and settling drop shapes at different values of $M$ (c) $D$ as a function of time (d) settling speed vs. $M$ plot for a PD-PD system with parameters $S=5$, $Ca=0.2$, $\lambda=1, \mathrm{Re}=15$.

### 6.2.2. Leaky-dielectric drop in a leaky-dielectric medium

Now we investigate the settling of a LD drop through a LD medium in the presence of electric field considering $\lambda = 1, Ca = 0.2$, $\text{Re} = 10$. Similar to Stokes regime, two combinations of electric properties ($R, S$) are chosen, such that for one case $\phi_T > 0$ and for another $\phi_T < 0$. Considering $(R, S) = (2.5, 1)$, the effect of Masson number ($M$) on drop deformation is shown in figure 9(a). For the settling drop, we compare the deformation for $Re_E = 0.01$ (negligible charge convection) and $Re_E = 0.8$ (significant charge convection) at a varying $M$. As $\phi_T > 0$, a freely suspended drop undergoes symmetric prolate deformation on application of electric field. For a LD drop settling at $\text{Re} = 10$, by increasing the value of $M$ (hence $M/\text{Re}$), resistance to oblate deformation becomes stronger analogous to the case of a PD drop. For $M = 3$, the drop becomes almost spherical ($D \sim -0.0085$). Further strengthening the electric field elongates the drop in the flow direction. Figure 9(a) suggests that the effect of $Re_E$ on drop deformation is quite weak. This is the reason for which the deformation values for $Re_E = 0.01$ and $Re_E = 0.8$ are alike. The steady-state deformation for $Re_E = 0.01$ is $D \sim 0.03485$ whereas for $Re_E = 0.8$ the value of deformation is $D \sim 0.03879$.

The variation of deformation with time is illustrated in figure 9(b) for $Re_E = 0.01$ and $Re_E = 0.8$ considering $M = 4$. The transient variation in the deformation is non-monotonic, similar to the case of the PD drop as described earlier in section 6.2.1. For $Re_E = 0.01$, charge relaxation time ($t_e = \varepsilon/\sigma$) is very less than flow time scale ($t_f = a/\bar{U}_{ref}$) whereas for $Re_E = 0.8$, both time scales are of same order. This is the reason why $t_{MD}$ (time corresponds to maximum deformation) for $Re_E = 0.80$ is more than that of $Re_E = 0.01$, as seen in figure 9(b). Initially, the drop elongates in flow direction up to $t = t_{MD}$, afterward the deformation decreases continuously to attain a steady-state value. When surface charge convection is taken into account, prolate drops deform more, accordingly for $Re_E = 0.8$ the drop shows extra deformation. Improved deformation in drop shape suggests weaker inertial stress relative to electric stress. The drop shape evolution is displayed in figure 9(c) for $M = 0, 2, 4$ considering small charge convection ($Re_E = 0.01$). Now we discuss the effect of Masson number ($M$) on drop settling speed. As discussed previously, for translating drops surface charge convection generates asymmetry in surface charge density and tangential electric stress, hence alters the motion of drop. For $(R, S) = (2.5, 1)$, surface charge convection resists the drop motion. With the increase in $M$ although overall settling speed increases due to decrease in cross-sectional area, however, the resistance to drop motion is getting better. Hence the deviation in settling speed increases with increase in $M$ in case of $Re_E = 0.8$ as shown in figure 9(d).

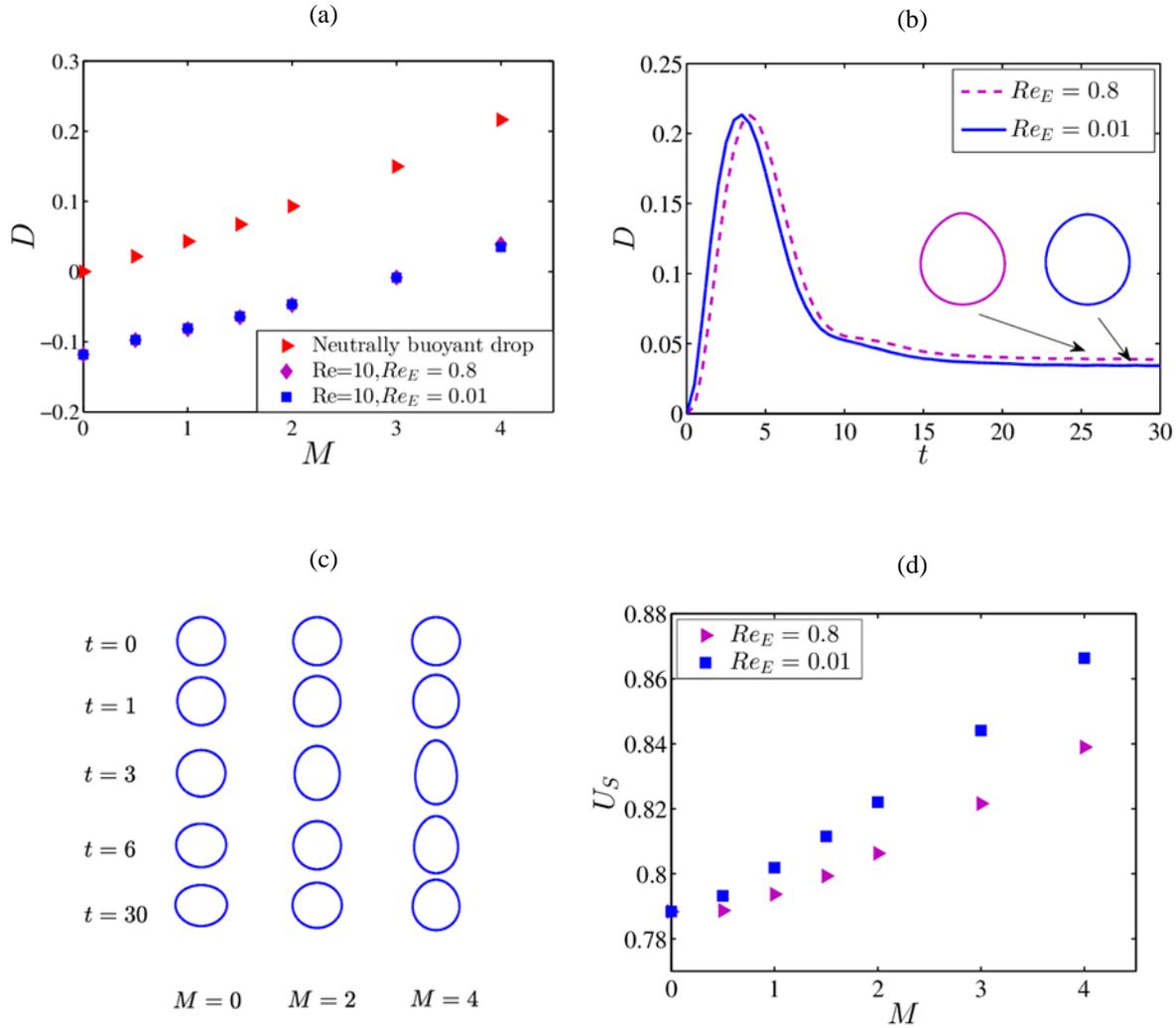

FIGURE 9. (a) Deformation ($D$) at different values of $M$ (b) $D$ as a function of time (c) transient drop shape evolution considering $Re_E = 0.01$ (d) settling speed vs $M$ plot for a LD-LD system. Other parameters are $R = 2.5, S = 1, Ca = 0.2, \lambda = 1, \text{Re} = 10$.

Now we consider another case where the drop undergoes oblate deformation for sole effect of the electric field ($\phi_T < 0$). Figure 10(a) shows the comparison between the effect of electric field on a neutrally buoyant and settling drop for $R = 1, S = 2.5, \lambda = 1, Ca = 0.2$. For the present choice of parameters, both the electric stress and inertia stress assist oblate deformation. For $Re_E = 0.80$ (significant charge convection) the deformation is less than that of $Re_E = 0.01$

(negligible charge convection). Unlike the case of $\phi_T > 0$, the deformation obtained for $Re_E = 0.8$ shows noticeable deviation from that of $Re_E = 0.01$. This departure become more

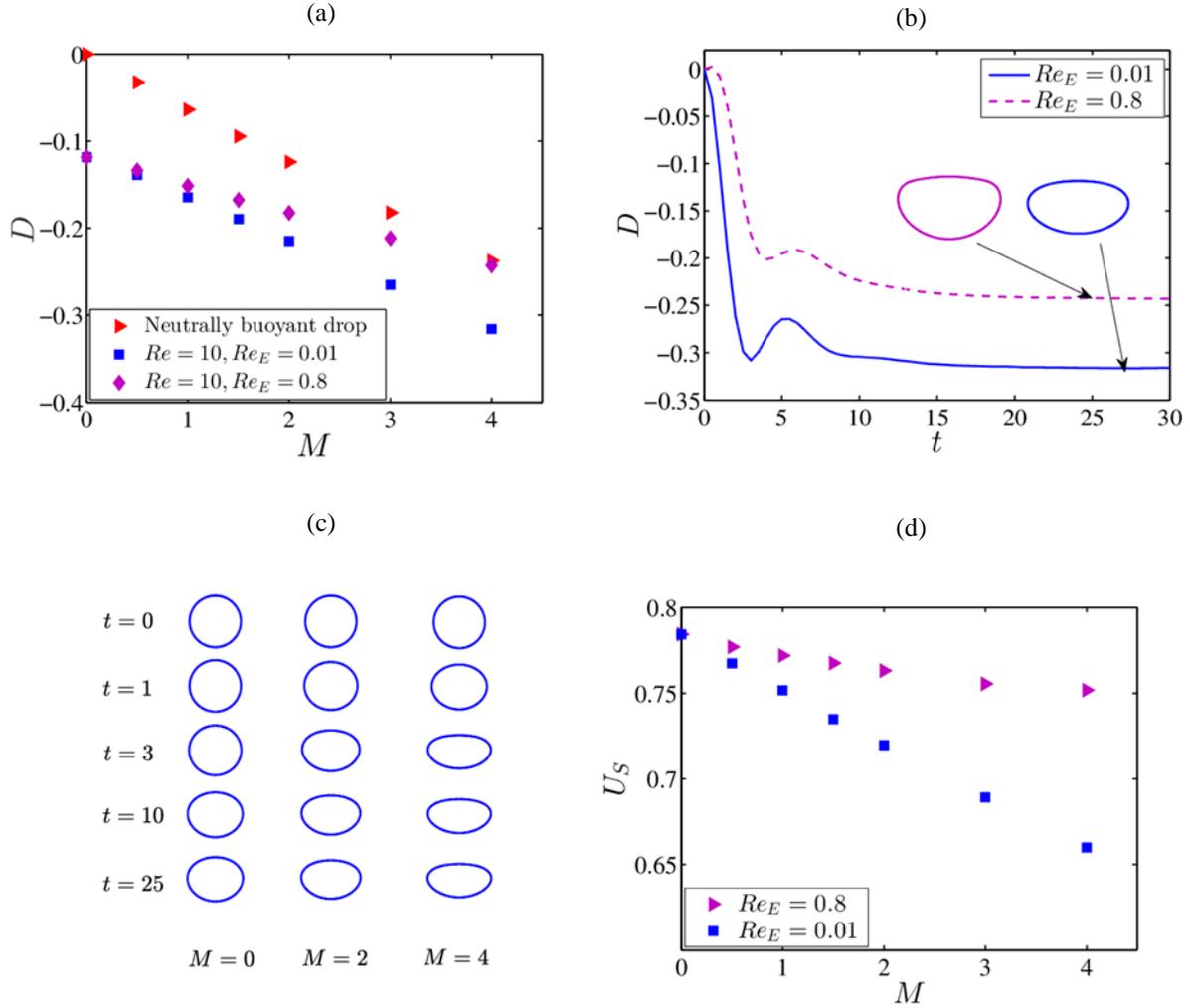

FIGURE 10. (a) Deformation ($D$) at different values of $M$ (b) $D$ as a function of time (c) transient drop shape evolution considering $Re_E = 0.01$ (d) settling speed vs. $M$ plot for a LD-LD system. Other parameters considered are $R = 1, S = 2.5, Ca = 0.2, \lambda = 1, \text{Re} = 10$.

significant at a higher value of $M$. At $M = 4$, the steady-state deformation obtained for $Re_E = 0.01$ is $D \sim -0.3160$ and for $Re_E = 0.80$ is $D \sim -0.2429$. As discussed in case of a prolate drop, charge convection weakens the relative strength of inertia stress. Owing to the similar phenomenon, oblate drop deforms less. The transient behavior of deformation is plotted in figure 10(b). For $Re_E = 0.01$, $t_e \ll t_f$, whereas in case of $Re_E = 0.80$, $t_e \sim t_f$. This results in slightly slow deformation in case of $Re_E = 0.8$. Effect of electric field strength on the time-dependent shape evolution is displayed in figure 10(c) of a settling drop for

$R=1, S=2.5, \lambda=1, Ca=0.2, Re_E=0.01, \text{Re}=10$. Figure 10(d) depicts the comparison between the drop speed for $Re_E=0.01$ and $Re_E=0.80$ at varying $M$. In absence of surface charge convection, the drop speed falls appreciably with an increase in Masson number ($M$) owing to increase in cross-sectional area. However, for significant charge convection ($Re_E=0.80$), the behavior of drop speed is different. For $(R,S)=(1,2.5)$, surface charge convection reduces the drag owing to asymmetric tangential electric stress distribution (discussed in section 6.1.2). Along with that drop deformation is quite less as that of negligible charge convection (figure 10a). The combination of above-mentioned effects, results in higher settling speed as seen in figure 10(d). An increase in Masson number intensifies the effect of charge convection, further improves the drop speed.

## 7. Conclusions

In the present work, EHD of settling drop is investigated under the action of uniform DC electric field applied parallel to the direction of gravitational field for Stokes regime and inertial regime. For Stokes regime, the EHD problem is solved analytically and numerically. Including the effect of significant electric field strength and surface charge convection, considering a leaky dielectric drop suspended in a leaky dielectric medium, we have performed asymptotic analysis in the limit of small capillary number $(Ca)$ to calculate the settling speed correct up to $O(Ca)$ and deformation correct up to $O(Ca^2)$. However, the numerical simulations are performed up to the higher value of $Ca$. In inertial regime, the effect of Masson number on transient drop dynamics is studied considering significant charge convection. Analyzing the results, we have concluded the present work as follows.

I. Electric field influenced deformation depends on the discriminating function $\phi_T(R,S,\lambda)$. For $\phi_T<0$, drop deforms into oblate shape and moves slower. When $\phi_T>0$, drop deforms into prolate shape and hence moves faster. In Stokes limit ($\text{Re}\to 0$), deformation and settling speed predicted by our theory agrees well with the present numerical results, shows noticeable improvement over asymptotic theory of Xu and Homsy (2006).

II. In Stokes limit, modification to drop speed owing to surface charge convection is a linear function $\chi(R,S)$, provided by most theories (Mandal *et al.* 2017; Xu & Homsy 2006) in the asymptotic limit. Drop moves faster for $\chi<0$ and vice versa. A similar nature of drop speed is observed in present theoretical analysis at small values of $Re_E$. We further extend our study to higher values of $Re_E$ and obtained the nonlinear variation in drop speed quite well, however our analytical results slightly departures from numerical results at a higher value of $Re_E$ as higher order multipoles are not taken into account.

III. The steady drop shape of a perfectly dielectric drop settling under the combined influence of uniform electric field and fluid inertia can be prolate or oblate depending on the value of Masson number. When the strength of electric field is small, the fluid inertia domiantes the drop shape and makes it oblate, and beyond the critical electric field strength the drop attains a steady-state prolate shape. The transient deformation history also shows prolate to oblate shape transition. Increase in Masson number always increases the drop speed.

IV. In inertial regime, for the leaky dielectric drop in a leaky dielectric medium, $\phi_T(R,S,\lambda)$ not really predicts the prolate or oblate deformation, rather signifies resistance or assistance to inertia induced oblate deformation. For $\phi_T > 0$, increase in Masson shows a transition from oblate to prolate shape deformation. A considerable charge convection ($Re_E \sim 1$), depicts a lesser drop speed, however the alternation in drop speed is quite significant than a similar case in Stokes flow. The prolate deformation is weakly affected by charge convection.

V. For $\phi_T < 0$, electric stress combined with inertia stress results in more oblate deformation and slower moving drop. Charge convection suggests a lesser deformed oblate drop. For $Re_E \sim 1$, the drop shows slower deformation and attains substantially less steady-state deformation than that of $Re_E = 0.01$. Lesser deformed drop alongside the asymmetric tangential stress greatly improves the drop speed. With the increase in electric field strength, the effect of charge convection gets intensified, hence drop deformation and settling speed deviates more.

**Appendix A. Effect of number of spherical harmonics on settling speed**

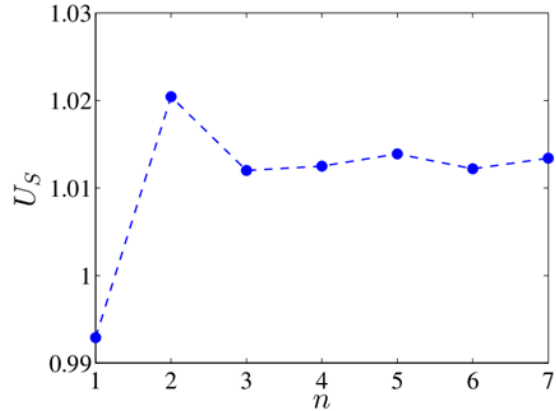

FIGURE A1. Variation of settling velocity with number of spherical harmonics ($n$) in leading-order for $R=1, S=2.5, \lambda=0.3571, Ca=0.05, M=1, Re_E=1$

Figure A1 describes the variation in settling speed $\left(U_S = U_S^{(0)} + CaU_S^{(Ca)}\right)$ with the number of spherical harmonics ($n$). The settling speed although oscillates with $n$, for $n \geq 4$ the deviation is negligible. To estimate the asymmetric deformation accurately and for ease of calculation, we have considered five spherical harmonics.

**Appendix B. Validation of Numerical procedure**

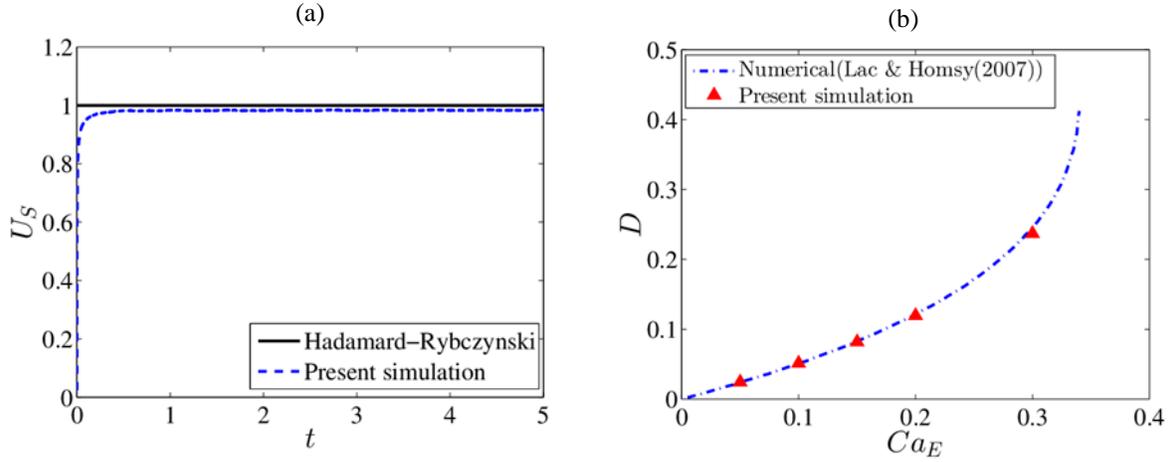

FIGURE B1. (a) Comparison of present numerical results with Hadamard-Rybczynski theory (b) comparison of deformation of a neutrally buoyant drop obtained by the present numerical method with numerical results of Lac and Homsy (2007). For case of neutrally buoyant drop $Re = 0.01$ is used.

Using a domain size of $70a \times 35a$, $\Delta z = W \times 2^{-11}$ grid size, in Stokes limit the velocity obtained by present numerical method agrees well with theoretical solution of Hadamard-Rybczynski as seen in figure B1 (a). For validation physical parameters considered are $\rho_r = 1.0406$, $\lambda = 0.3571$, $Re = 0.0024$.

To validate our numerical procedure for EHD problem, we compare the electric field induced deformation obtained from present numerical results with that of Lac and Homsy (2007) illustrated by figure B1(b) considering $R = 10$, $S = 1$, $\lambda = 1$, $Re = 0.01$, $Re_E = 0.01$. From figure B1(b), it is evident that shows that our numerical results matches very well with the numerical solution of Lac and Homsy (2007). Domain size of $20a \times 20a$ and grid size of $\Delta z = W \times 2^{-11}$ is used.

**Appendix C. Grid and Domain Independence test**

In figure C1(a) drop shapes with the vertical position is shown at time $t = 10$ for 3 different grid sizes $\Delta z = W \times 2^{-12}$, $\Delta z = W \times 2^{-11}$, $\Delta z = W \times 2^{-10}$ at the interface keeping bulk grid size of

$\Delta z = W \times 2^{-6}$. A computational domain of $60a \times 20a$ is used. Looking to figure C1(a) it can be said that the present numerical setup achieved grid convergence. To conduct domain independence test, two different domain size of $60a \times 20a$ and $75a \times 25a$ are considered. The deformed drop shape and vertical position is compared at $t = 20$ in figure C1(b) using a grid size of $\Delta z = W \times 2^{-11}$ near the interface. From the figure C1(b) it is clear that increasing the size of the computational domain puts negligible effect on drop dynamics.

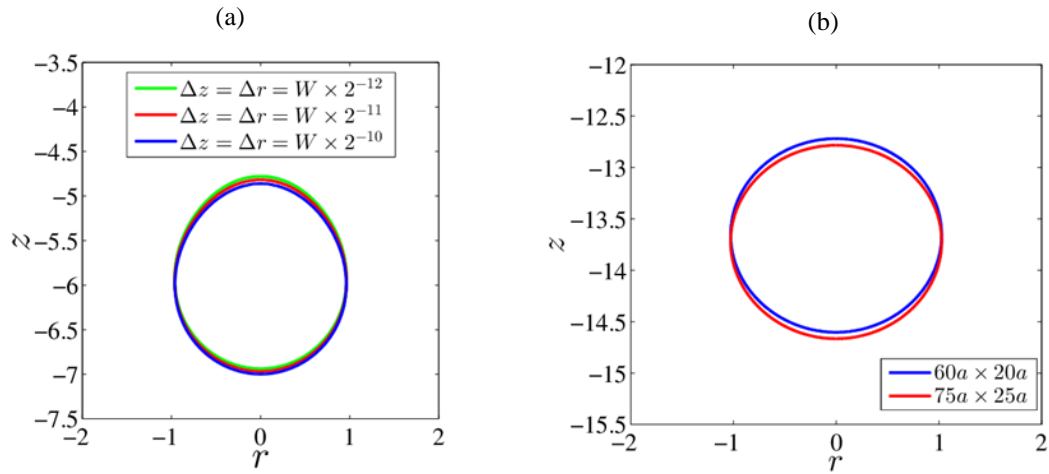

FIGURE C1. (a) Effect of grid size on numerical solution (b) effect of domain size on drop dynamics for $R = 2.5, S = 1, \lambda = 1, Ca = 0.2, \mathrm{Re} = 10$. Other parameters considered are $Re_E = 0.8$ and $M = 4$.